\documentclass[iop,apj,appendixfloats]{emulateapj}
\usepackage{graphicx}
\usepackage{amsmath}
\usepackage{amssymb}
\usepackage{color}

\gdef\xx[#1]{\textcolor{red}{#1}}

\gdef\kms{km\,s$^{-1}$}
\gdef\msun{M$_{\odot}$}

\gdef\blob{NGC1052--DF2}
\gdef\natpap{vD18}

\lefthead{van Dokkum et al.}
\righthead{}
\slugcomment{Accepted for publication in ApJ Letters}
\begin{document}

\title{
An Enigmatic Population of Luminous Globular Clusters in
a Galaxy Lacking Dark Matter
}

\author{Pieter van Dokkum\altaffilmark{1},
Yotam Cohen\altaffilmark{1},
Shany Danieli\altaffilmark{1},
J.~M.~Diederik Kruijssen\altaffilmark{2},
Aaron J.\ Romanowsky\altaffilmark{3,4},
Allison Merritt\altaffilmark{5},
Roberto Abraham\altaffilmark{6},
Jean Brodie\altaffilmark{3},
Charlie Conroy\altaffilmark{7},
Deborah Lokhorst\altaffilmark{6},
Lamiya Mowla\altaffilmark{1}, 
Ewan O'Sullivan\altaffilmark{7},
Jielai Zhang\altaffilmark{6}
\vspace{8pt}}

\altaffiltext{1}
{Astronomy Department, Yale University, 52 Hillhouse Ave,
New Haven, CT 06511, USA}
\altaffiltext{2}
{Astronomisches Rechen-Institut, Zentrum f\"ur Astronomie der Universit\"at
Heidelberg, M\"onchhofstra{\ss}e 12-14, D-69120 Heidelberg, Germany
}
\altaffiltext{3}
{University of California Observatories, 1156 High Street, Santa
Cruz, CA 95064, USA}
\altaffiltext{4}
{Department of Physics and Astronomy, San Jos\'e State University,
San Jose, CA 95192, USA}
\altaffiltext{5}
{Max-Planck-Institut f\"ur Astronomie, K\"onigstuhl 17, D-69117
Heidelberg, Germany}
\altaffiltext{6}
{Department of Astronomy \& Astrophysics, University of Toronto,
   50 St.\ George Street, Toronto, ON M5S 3H4, Canada}
\altaffiltext{7}
{Harvard-Smithsonian Center for Astrophysics, 60 Garden Street,
Cambridge, MA, USA}

\begin{abstract}

We recently found an ultra diffuse
galaxy (UDG) with a half-light radius of $R_e=2.2$\,kpc
and little or no dark matter. The total mass
of  \blob\ was measured from the
radial velocities of bright compact objects that are associated
with the galaxy.
Here we analyze these objects using
a combination of {\em HST} imaging and Keck
spectroscopy.
%Their luminosity function has a narrow peak at
%$M_{V,606} = -9.1$, a factor of
%$\sim 4$ brighter than the canonical peak of the globular cluster
%luminosity function. 
%The combined luminosity of the 11 confirmed clusters is
%4\,\% of the total luminosity of \blob.
Their average size is $\langle{}r_h\rangle=6.2\pm
0.5$\,pc and their average ellipticity is
$\langle\epsilon\rangle=0.18\pm 0.02$.
%a factor of 2.2 larger than
%typical Milky Way glubular clusters.
%They are also more elongated, with an average
%ellipticity of 
%The stellar populations
%of the clusters are remarkably uniform:
From a stacked Keck spectrum we derive an
age of $\gtrsim 9$ Gyr and a metallicity of
${\rm [Fe/H]}=-1.35\pm{}0.12$.
%In terms of their luminosity, structure, and stellar populations
%the 11 spectroscopically-confirmed compact
%objects
Their properties are similar to $\omega$\,Centauri,
the brightest and largest globular cluster in the Milky Way, and
%It is not clear how these compact objects were formed, and
%what their connection is with the diffuse light
%that constitutes the rest of the galaxy.
our results demonstrate that the luminosity function
of metal-poor
globular clusters is not universal.
%The observed scatter in their
%$V_{606}-I_{814}$ colors is only 0.039\,mag, and their mean color
%is identical within the errors to that of the diffuse light.
The fraction of the total stellar mass
that is in the globular cluster system is similar
to that in other UDGs, and consistent with
``failed galaxy'' scenarios where star formation terminated shortly
after the clusters were formed. However, the galaxy
is a factor of $\sim{}1000$ removed from the 
relation between globular cluster
mass and total galaxy mass that has been found for other galaxies,
including other UDGs.
We infer that a dark matter halo is not
a prerequisite for the formation of metal-poor globular cluster-like
objects in high redshift galaxies.
%of dark matter.
%this relation is a by-product of th
%to form luminous globular clusters in the absence of dark matter.
%It may be that star formation proceeds differently
%in the absence of dark matter, or that the event that shut off
%star formation in NGC1052-DF2
%also removed most of its dark matter. Regardless of their origin,
%the clusters cannot have experienced significant
%tidally-induced mass loss since their formation,
%and their luminosity function may be relatively pristine.

\end{abstract}


\section{Introduction}

\hyphenation{kruijssen}

%not strange in itself
%unlike other udgs - more luminous
% enables good constraints on stellar populations
% also different in mass - gc relation; implications for formation

We recently identified a galaxy
with little or no dark matter ({van Dokkum} {et~al.} 2018, hereafter \natpap).
\blob\ has a stellar mass of $M_{\rm stars}\approx 2
\times 10^8$\,\msun\ and a 90\,\% confidence upper limit on its dark
matter halo mass of $M_{\rm halo}<1.5\times 10^8$\,\msun, placing it
a factor of $\gtrsim 400$ off of the canonical stellar mass -- halo
mass relation (Moster et al.\ 2013; {Behroozi} {et~al.} 2013).
\blob\ is a featureless, spheroidal
``ultra diffuse'' galaxy (UDG; {van Dokkum} {et~al.} 2015),
with an effective radius of $R_e = 2.2$\,kpc and a central
surface brightness $\mu(V_{606},0) = 24.4$\,mag\,arcsec$^{-2}$.
It has a radial velocity of $1803$\,\kms. Its
SBF-determined distance is $19.0\pm 1.7$\,Mpc (\natpap), consistent with
that of the NGC\,1052 group at $D\approx 20$\,Mpc ({Blakeslee} {et~al.} 2010).

The kinematics of \blob\ were measured from the radial velocities of 10
compact objects
that are associated with the galaxy.
These  objects drew our attention to the galaxy in
the first place: it is a large, low surface brightness blob in our
Dragonfly Telephoto Array imaging ({Abraham} \& {van Dokkum} 2014; {Merritt} {et~al.} 2016) but a
collection of point-like sources in the Sloan Digital Sky Survey.
%(SDSS;  ).

Finding globular clusters (GCs) in a UDG is
in itself not unusual
({Beasley} {et~al.} 2016; {Peng} \& {Lim} 2016; {van Dokkum} {et~al.} 2016, 2017; {Amorisco}, {Monachesi}, \&  {White} 2018). 
In fact, Coma UDGs have on average $\sim 7$\
times more GCs than other galaxies of the same
luminosity ({van Dokkum} {et~al.} 2017), with large galaxy-to-galaxy scatter
({Amorisco} {et~al.} 2018).
However, what {\em is} unusual, or at least unexpected, is the
remarkable luminosity of the clusters. 
The luminosity function of the GC
populations of Coma UDGs is consistent with that seen in other galaxies,
peaking at an absolute magnitude
$M_V \sim -7.5$ ({Peng} \& {Lim} 2016; {van Dokkum} {et~al.} 2017; {Amorisco} {et~al.} 2018). The
ten clusters that
were analyzed in \natpap\ are all significantly brighter than this,
raising the question whether the GC luminosity function is
systematically offset from that in other galaxies.

Here we focus on the properties of the compact objects in
\blob, using imaging from the
{\em Hubble Space Telescope} ({\em HST})  and spectroscopy
obtained with the W.~M.~Keck Observatory.
%We first ask whether the spectroscopically-confirmed clusters
%are the bright end of a ``normal'' luminosity function, or the
%GC luminosity function is shifted to brighter magnitudes in \blob.
%Next, we analyze the structure and stellar populations of
%the clusters.
We show that the GC system of \blob\ is unprecedented,
both in terms of the average properties of the clusters and in its offset
from the canonical scaling relation between GC system mass and
total galaxy mass.

\begin{figure*}[htbp]
  \begin{center}
  \includegraphics[width=0.9\linewidth]{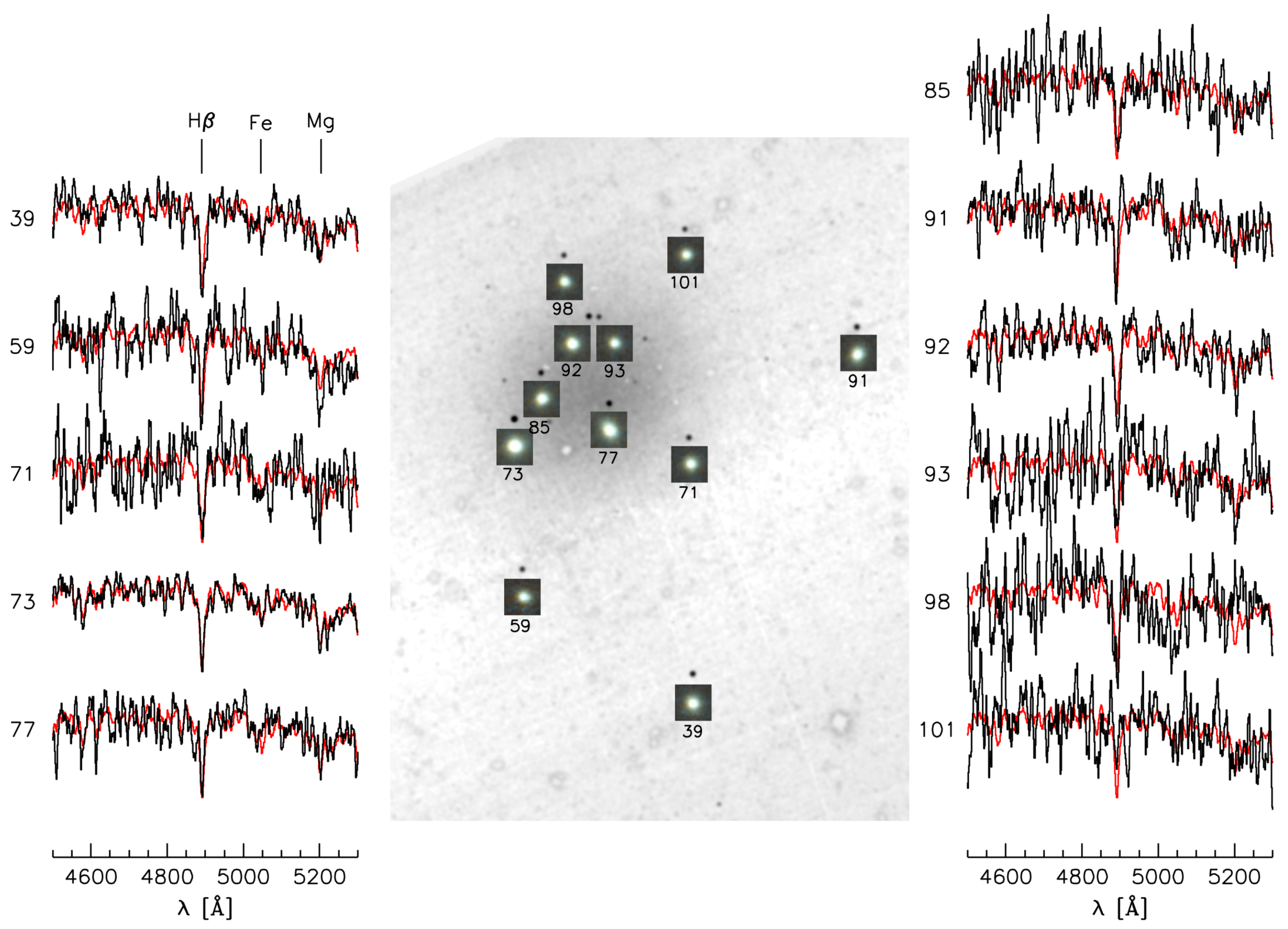}
  \end{center}
\vspace{-0.2cm}
    \caption{
Keck/LRIS spectra (left and right) and {\em HST} images (center)
of the 11 clusters
associated with \blob. The color images, generated
from the $V_{606}$ and $I_{814}$ data, span $1\arcsec \times
1\arcsec$. Some of the clusters are visibly flattened. 
The background image was generated by masking all objects in 
the $I_{814}$ {\em HST} frame that do
not match the color and size criteria we use for selecting GCs,
and then applying a slight smoothing to emphasize the compact
objects. %A wider view is shown in Fig.\ \ref{selglobs.fig}.
The spectra focus on the
wavelength region around the redshifted $\lambda{}4861$\,H$\beta$
and $\lambda{}5172$\,Mg lines. The red line is a S/N-weighted average
of the 11 spectra.
}
\label{spectra.fig}
\end{figure*}

\section{Identification}

\subsection{Spectroscopically-Identified Clusters}

We obtained spectra of compact objects in the \blob\ region with
the Keck telescopes, using
the Deep Imaging Multi-Object Spectrograph 
on Keck II, the red arm of the Low-Resolution Imaging
Spectrometer (LRIS; {Oke} {et~al.} 1995), and the blue arm of
LRIS. The sample selection, reduction, and analysis of the
high resolution DEIMOS and red LRIS data are described in detail
in \natpap. The 
blue-side LRIS data
were obtained with the 300/5000 grism
and $1\farcs0$ slits,
providing a spectral resolution ranging from $\sigma_{\rm instr}
\sim{}350$\,\kms\ at $\lambda=3800$\,\AA\ to $\sigma_{\rm
instr}\sim{}150$\,\kms\
at $\lambda=6600$\,\AA. The reduction followed the same procedures as
the red side data, and is described in \natpap.
The spectral resolution
is too low for accurate radial velocity measurements, but the wide
wavelength coverage provides
constraints on the stellar populations (\S\,\ref{sps.sec}).
Small sections of the
spectra of the 11 confirmed GCs are shown in Fig.\ \ref{spectra.fig}.
Note that we analyze one more object in this paper than in \natpap;
this is because
the S/N ratio of the red spectrum of GC-93 is too low for
an accurate velocity measurement.\footnote{Oddly the red side spectrum of
GC-93 appears to be
featureless in the Ca\,triplet region.
%This may simply be a result
%of the low S/N ratio, but it would be interesting to obtain a deeper
%red spectrum of this object.
}

\begin{figure*}[htbp]
  \begin{center}
  \includegraphics[width=0.95\linewidth]{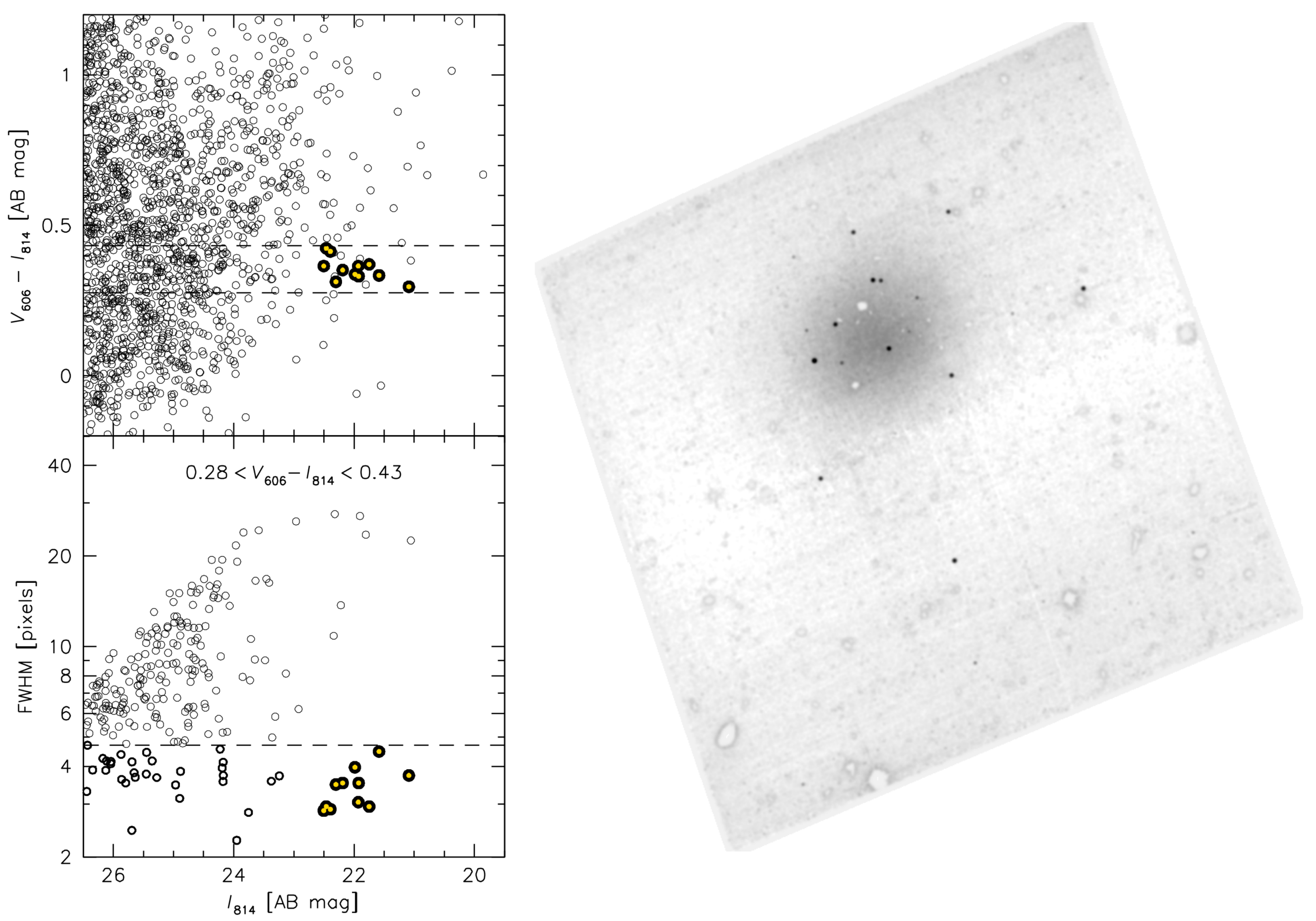}
  \end{center}
\vspace{-0.2cm}
    \caption{
Photometric selection of globular clusters. The top panel shows the
color-magnitude relation of all
objects in the {\em HST} images of \blob. The 11 spectroscopically-confirmed objects
are marked with yellow and black circles. Dashed lines delineate  the
$\pm{}2\sigma$ range of the colors of the confirmed clusters:
$0.28<V_{606}-I_{814}<0.43$. The bottom panel shows the size-magnitude
relation for all objects that satisfy this color criterion. Objects with
${\rm FWHM}<4.7$\,pixels are candidate GCs.
%The spectroscopic
%completeness is 100\,\% for $I_{814}<23$. 
The image on the right is a wider view of
that shown in Fig.\ \ref{spectra.fig}. All objects
are masked, except those that match the color and size criteria.
}
\label{selglobs.fig}
\end{figure*}

\subsection{Photometrically-Identified Clusters}

%The Keck masks include the majority of bright, compact objects in the
%vicinity of \blob. However, 
To measure  the luminosity function we also have to consider
GCs that are fainter
than the spectroscopic limits, as well as any that might not
have been included in the masks. We select all candidate GCs using
the $V_{606}$ and $I_{814}$ {\em HST} images
(described in \natpap).
Photometric catalogs were created using SExtractor ({Bertin} \& {Arnouts} 1996) in
dual image mode. The photometry was corrected for Galactic
extinction (Schlafley \& Finkbeiner 2011),
and the $V_{606}-I_{814}$ colors were corrected for
the wavelength dependence of the PSF. Total magnitudes were determined
from the ``AUTO'' fluxes, with an object-by-object
correction to an infinite aperture
as determined from
the encircled energy curves of {Bohlin} (2016).

The top panel in Fig.\ \ref{selglobs.fig} shows all objects with
$I_{814}<26.5$ in the plane of $V_{606}-I_{814}$ color vs.\ $I_{814}$
magnitude. The 11 spectroscopically-identified clusters have a
remarkably small range in color: we find
$\langle{}V_{606}-I_{814}\rangle=0.36$ with an observed rms scatter of
$\sigma_{V-I}=0.039$. This is not a result of selection;
we obtained spectra of nearly all compact objects in the vicinity
of \blob\ irrespective of their color. The bottom panel of
Fig.\ \ref{selglobs.fig} shows the relation between the SExtractor
FWHM and $I_{814}$ magnitude for all objects that have colors
in the range $\langle{}V_{606}-I_{814}\rangle\pm2\sigma_{V-I}$.
We note that the results are not sensitive to the precise limits that
are used here.
As expected, the spectroscopically-identified
GCs are small. The dashed line corresponds to
${\rm FWHM}<\langle{\rm FWHM}\rangle+2.5\sigma_{\rm FWHM} = 4.7$\,pixels.

We find that the spectroscopic completeness is 100\,\% for $I_{814}<23$
objects that satisfy the color and size criteria. We find 16 candidate
GCs with $23<I_{814}<25.5$, but as we show below most
are probably compact background galaxies.
The grey scale panel of Fig.\ \ref{selglobs.fig}
shows the $I_{814}$ data after masking
all objects that do {\em not} satisfy these
criteria. The masked image was smoothed with a Gaussian of
${\rm FWHM}=0\farcs 9$.
%The bright clusters are clearly associated with the low surface brightness
%emission. The contrast between the GCs and the smooth appearance of
%\blob\ is striking.

\begin{figure*}[htbp]
  \begin{center}
  \includegraphics[width=0.7\linewidth]{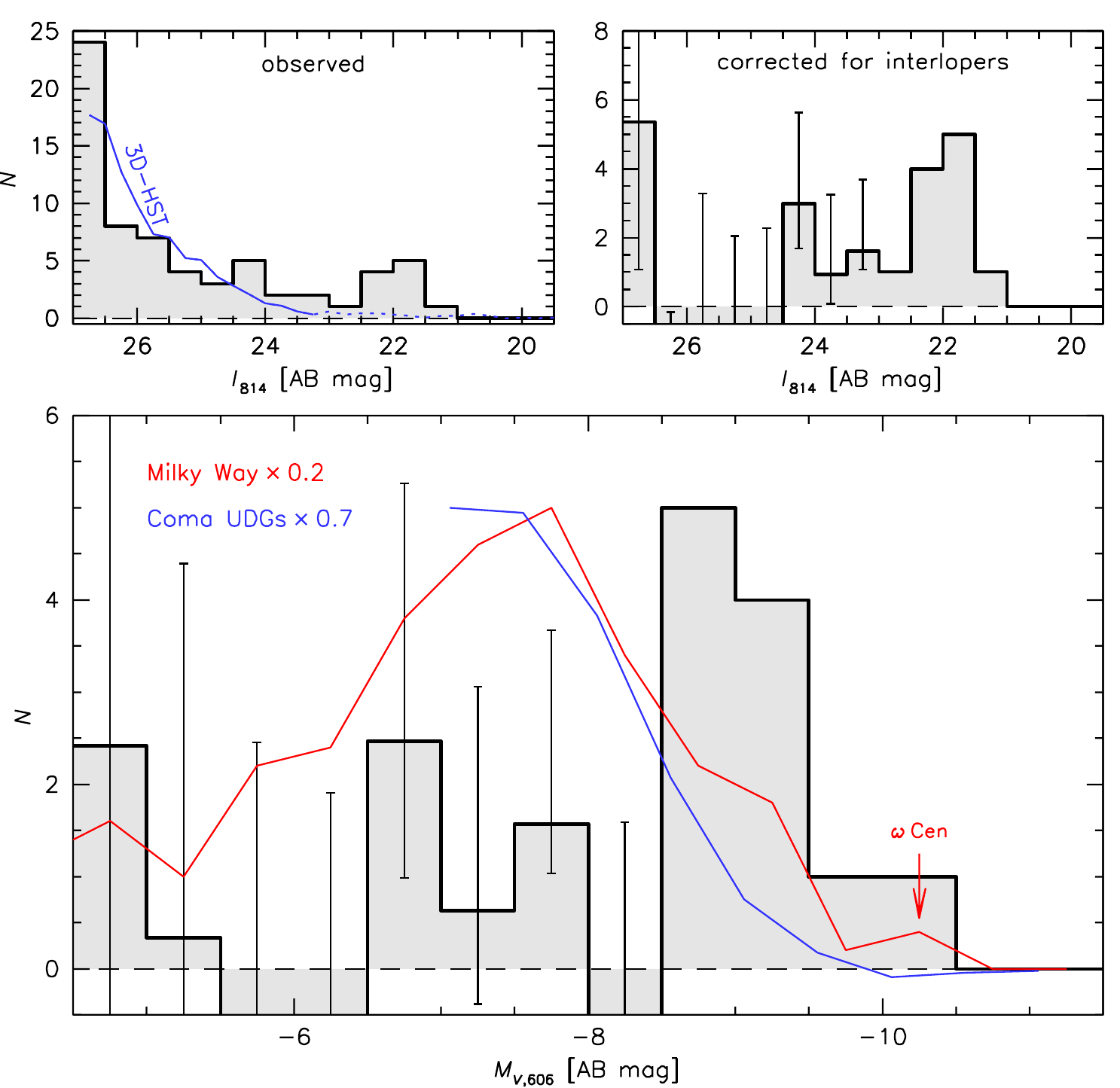}
  \end{center}
\vspace{-0.2cm}
    \caption{
Luminosity function of the compact objects in \blob.
{\em Top left:} Observed luminosity function,
in apparent $I_{814}$ magnitude. The blue line shows the magnitude
distribution of objects in blank field
3D-HST/CANDELS imaging that have the same colors and sizes as
the GCs. {\em Top right:} Observed luminosity function, after
correcting each bin for the expected number of unrelated objects.
%The luminosity function has a narrow peak at $I_{\rm 814} \approx 22$.
{\em Bottom:} Luminosity function in absolute magnitude, for $D=20$\,Mpc.
The luminosity functions of GCs in the Milky
Way and in Coma UDGs are shown in red and blue, respectively.
%{\em Bottom:} Distribution of the kinetic energy density
%$E/V = P \propto M^2r_h^{-4}$, normalized to the median for the Milky
%Way. This corresponds roughly to
%$P\sim 10^8k$\,cm$^{-3}$\,K ( ).
}
\label{lf.fig}
\end{figure*}

\section{Luminosity Function and Specific Frequency}
\label{lf.sec}
At bright magnitudes it is straightforward to measure the luminosity
function of the GCs because the spectroscopic completeness is 100\,\%,
but at $I_{814}>23$ a correction needs to be made for interlopers.
This is evident from the
distribution of objects in the bottom panel of Fig.\
\ref{selglobs.fig}: at $I_{814}<23$ the GCs are well-separated from
other objects, but at faint magnitudes there is a continuous distribution
of sources with ${\rm FWHM}\sim 2-15$\,pixels.
This magnitude-dependent correction for unrelated objects was determined from
ACS imaging obtained in the blank field CANDELS survey ({Koekemoer} {et~al.} 2011).
We obtained CANDELS $V_{606}$ and $I_{814}$ images of the AEGIS field from
the 3D-HST data release ({Skelton} {et~al.} 2014), and analyzed these in the exact
same way as the \blob\ data.
% (taking the small difference in pixel
%scale\footnote{The CANDELS data are sampled at $0\farcs 06$\,pix$^{-1}$ and
%ours at $0\farcs 05$\,pix$^{-1}$.} into account).

The results are shown in the top panels of Fig.\ \ref{lf.fig}.
The expected contamination increases
steadily with magnitude at $I_{814}>23$.
%distribution as the objects in \blob.
The top right panel shows the
observed magnitude distribution after subtracting the expected contamination,
with the uncertainties reflecting the Poisson errors in the observed
counts in each bin. There is a pronounced peak at $I_{814}=22.0$
with a $1\sigma$ width of 0.4\,mag, consisting of the 11 confirmed
clusters.

The bottom panel of Fig.\ \ref{lf.fig} shows the luminosity function.
For consistency with other work we focus on $M_{V,606}$, determined
from the total $I_{814}$ magnitudes through $M_{V,606}=I_{814}
+(V_{606}-I_{814})-31.50$.
The mean absolute magnitude of the confirmed clusters is $M_{V,606}=-9.1$,
and the brightest cluster (GC-73) has $M_{V,606}=-10.1$.
The red curve shows the (scaled) luminosity function of Milky Way GCs,
obtained from the 2010 edition of the {Harris} (1996)
catalog\footnote{http://physwww.mcmaster.ca/\~{ }harris/mwgc.dat}
with $M_{V,606}=M_V-0.05$.
The peak magnitude of $M_V\sim{}-7.5$
for the Milky Way is similar to that seen in other galaxies
(e.g., {Rejkuba} 2012).
The blue curve is the average luminosity function of GCs in
the two UDGs Dragonfly~44 and DFX1, taken from {van Dokkum} {et~al.} (2017).

The luminosity function of \blob\ is shifted to higher
luminosities than those of other galaxies, including other
UDGs. The difference is a factor of $\sim 4$. Phrased differently,
the GC luminosity function of \blob\ is not far removed from the
bright end of the luminosity function of the Milky Way:
\blob\ has
11 clusters brighter than $M_{V,606}=-8.6$, whereas the Milky
Way has 20 
(and only 15 with ${\rm [Fe/H]}<-1$).
However,
there is only marginal evidence for the presence of ``classical''
GCs with $M_{V,606}\sim -7.5$ 
in \blob: after correcting for
interlopers, the total number of GCs with $-8.5<M_{V,606}<-6.5$
is $N_{\rm peak}=4.2^{+3.4}_{-2.1}$ (compared to $N_{\rm peak}=84$
in the Milky Way).

Taking the total number of globular clusters as $\approx 15$,
we derive a specific frequency 
$S_N\equiv{}N_{\rm GC}\times 10^{0.4 (M_V^{\rm g}+15)}\approx{}11$,
where $M_V^{\rm g}=-15.4$ is the total magnitude of the
galaxy (see \natpap).
The 11 spectroscopically-confirmed clusters
constitute 4\,\% of the total luminosity
of \blob\ (with 1\,\% contributed by GC-73 and 3\,\% by the other
clusters).

\section{Structural Parameters}

%Globular clusters have characteristic half-light radii $r_h\sim 3$\,pc
%and are nearly round
%, with an average ellipticity $\langle\epsilon
%\rangle\equiv 1-\langle{}b/a\rangle{}\sim
%0.07$
%(e.g., {Harris} 1996;  ).
%The \blob\ GCs
%are resolved in the {\em HST} data, and here
We use the {\em HST} imaging
to  compare the morphologies of the \blob\ GCs to
those of Milky Way GCs. We fit {King} (1962) models
% (with $\alpha=2$)
to the
individual {\tt .flc} files 
using the GALFIT software ({Peng} {et~al.} 2002) with
synthetic PSFs.
This provides eight independent measurements (four in $V_{606}$ and
four in $I_{814}$). Cosmic rays
%(identified with L.A.\,Cosmic;  ) 
and
neighboring objects were masked in the fits.

The results are listed in Table 1.
Circularized
half-light radii $r_h$ were determined from the measured core and tidal
radii (multiplied by $\sqrt{b/a}$).
The listed values are the
biweight averages (see {Beers}, {Flynn}, \& {Gebhardt} 1990)
of the eight individual measurements, and for each entry
the listed error is the biweight
scatter in the eight individual measurements.
%\footnote{This is pessimistic; the
%formal error on the mean is $\approx s/\sqrt{8}$ with $s$ the listed error.}
We verified that very similar values are obtained if a {Sersic} (1968)
profile is fitted to the objects instead of a King profile.
As a test of
our ability to measure the sizes of these small objects
we also included
four stars of similar brightness to the GCs in the fits.
All four stars
have $r_h<0\farcs{}018$, whereas the GCs have sizes in the
range $0\farcs{}043\leq r_h\leq 0\farcs{}089$.

\begin{figure}[htbp]
  \begin{center}
  \includegraphics[width=0.95\linewidth]{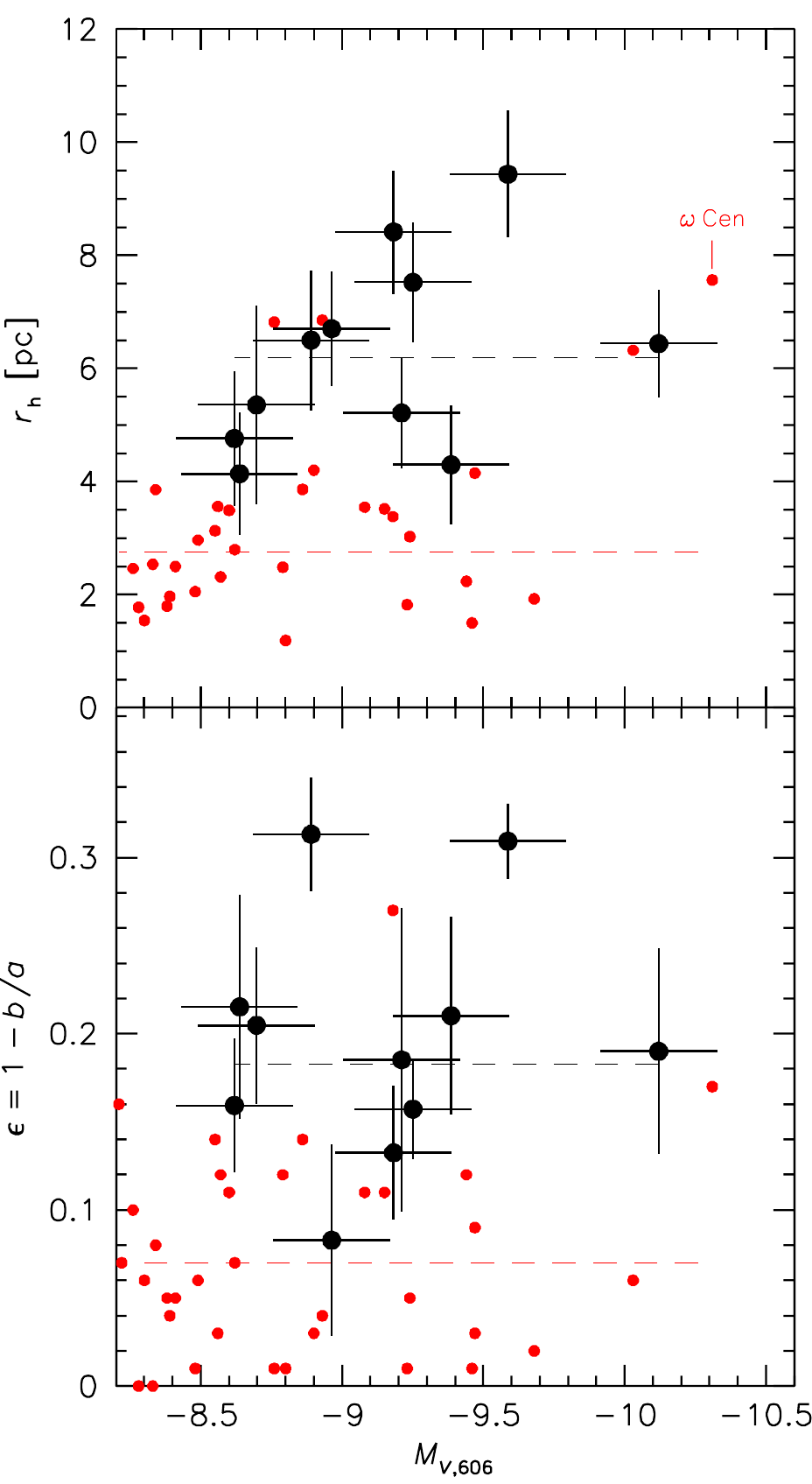}
  \end{center}
\vspace{-0.2cm}
    \caption{
Morphological parameters of the GCs. The top panel shows the
circularized half-light
radii versus the absolute magnitude, for \blob\ (black points with
error bars) and the Milky Way (red). Errors in $M_{V,606}$
and $r_h$
include a 10\,\% uncertainty in the distance (see \natpap).
The bottom panel shows
the ellipticity. Means are indicated with dashed lines.
%The 11 GCs in \blob\ have a mean size of
%$\langle{}r_h\rangle=6.2$\,pc and a mean ellipticity of
%$\langle\epsilon\rangle=0.18$ (indicated with black dashed lines).
%These values are higher than those of Milky Way clusters in the
%same luminosity range (red dashed lines).
}
\label{structure.fig}
\end{figure}

The sizes and ellipticities are compared to those of Milky Way GCs
in Fig.\ \ref{structure.fig}, again making use of the 2010
version of the {Harris} (1996) compilation. The (biweight)
mean size of the 11 objects is $\langle{}r_h\rangle=6.2\pm 0.5$\,pc, a factor
of 2.2 larger than the mean size of Milky Way GCs
in the same luminosity range. The mean ellipticity is
$\langle\epsilon\rangle=0.18\pm 0.02$, a factor of 2.6 larger than
Milky Way GCs.
%We conclude that the structure of the 11 compact objects
%is, on average, different from that of ``normal'' GCs: they are
%significantly larger and more flattened. 

\noindent
\begin{deluxetable}{cccccc}
\tablecaption{Properties Of Globular Clusters\label{prop.tab}}
\tabletypesize{\footnotesize}
\tablehead{\colhead{Id} & \colhead{RA} & \colhead{DEC}
 & \colhead{$M_{V,606}$} &
\colhead{$r_h$\tablenotemark{a}}
& \colhead{$\epsilon$}}
\startdata
39 & 2$^{\rm h}41^{\rm m}45.07^{\rm s}$ & $-8\arcdeg25\arcmin24\farcs9$
& $-9.3$ & $7.5\pm0.7$ & $0.16\pm0.03$\\
59 & 2$^{\rm h}41^{\rm m}48.08^{\rm s}$ & $-8\arcdeg24\arcmin57\farcs5$
& $-8.9$ & $6.5\pm1.0$ & $0.31\pm0.03$\\
71 & 2$^{\rm h}41^{\rm m}45.13^{\rm s}$ & $-8\arcdeg24\arcmin23\farcs0$
& $-9.0$ & $6.7\pm0.8$ & $0.08\pm0.05$\\
73 & 2$^{\rm h}41^{\rm m}48.22^{\rm s}$ & $-8\arcdeg24\arcmin18\farcs1$
& $-10.1$ & $6.4\pm0.7$ & $0.19\pm0.06$\\
77 & 2$^{\rm h}41^{\rm m}46.54^{\rm s}$ & $-8\arcdeg24\arcmin14\farcs0$
& $-9.6$ & $9.4\pm0.6$ & $0.31\pm0.02$\\
85 & 2$^{\rm h}41^{\rm m}47.75^{\rm s}$ & $-8\arcdeg24\arcmin05\farcs9$
& $-9.2$ & $5.2\pm0.8$ & $0.19\pm0.09$\\
91 & 2$^{\rm h}41^{\rm m}42.17^{\rm s}$ & $-8\arcdeg23\arcmin54\farcs0$
& $-9.2$ & $8.4\pm0.7$ & $0.13\pm0.04$\\
93 & 2$^{\rm h}41^{\rm m}46.72^{\rm s}$ & $-8\arcdeg23\arcmin51\farcs3$
& $-8.6$ & $4.1\pm1.0$ & $0.22\pm0.06$\\
92 & 2$^{\rm h}41^{\rm m}46.90^{\rm s}$ & $-8\arcdeg23\arcmin51\farcs1$
& $-9.4$ & $4.3\pm1.0$ & $0.21\pm0.06$\\
98 & 2$^{\rm h}41^{\rm m}47.34^{\rm s}$ & $-8\arcdeg23\arcmin35\farcs2$
& $-8.7$ & $5.4\pm1.7$ & $0.20\pm0.04$\\
101 & 2$^{\rm h}41^{\rm m}45.21^{\rm s}$ & $-8\arcdeg23\arcmin28\farcs3$
& $-8.6$ & $4.8\pm1.1$ & $0.16\pm0.04$
\enddata
\tablenotetext{a}{Circularized half-light radius of King profile,
in parsecs.
%The listed errors do not include the $\sim 10$\,\% uncertainty
%in the distance of \blob.
}
\end{deluxetable}

\section{Stellar Populations}
\label{sps.sec}

\begin{figure*}[htbp]
  \begin{center}
  \includegraphics[width=0.8\linewidth]{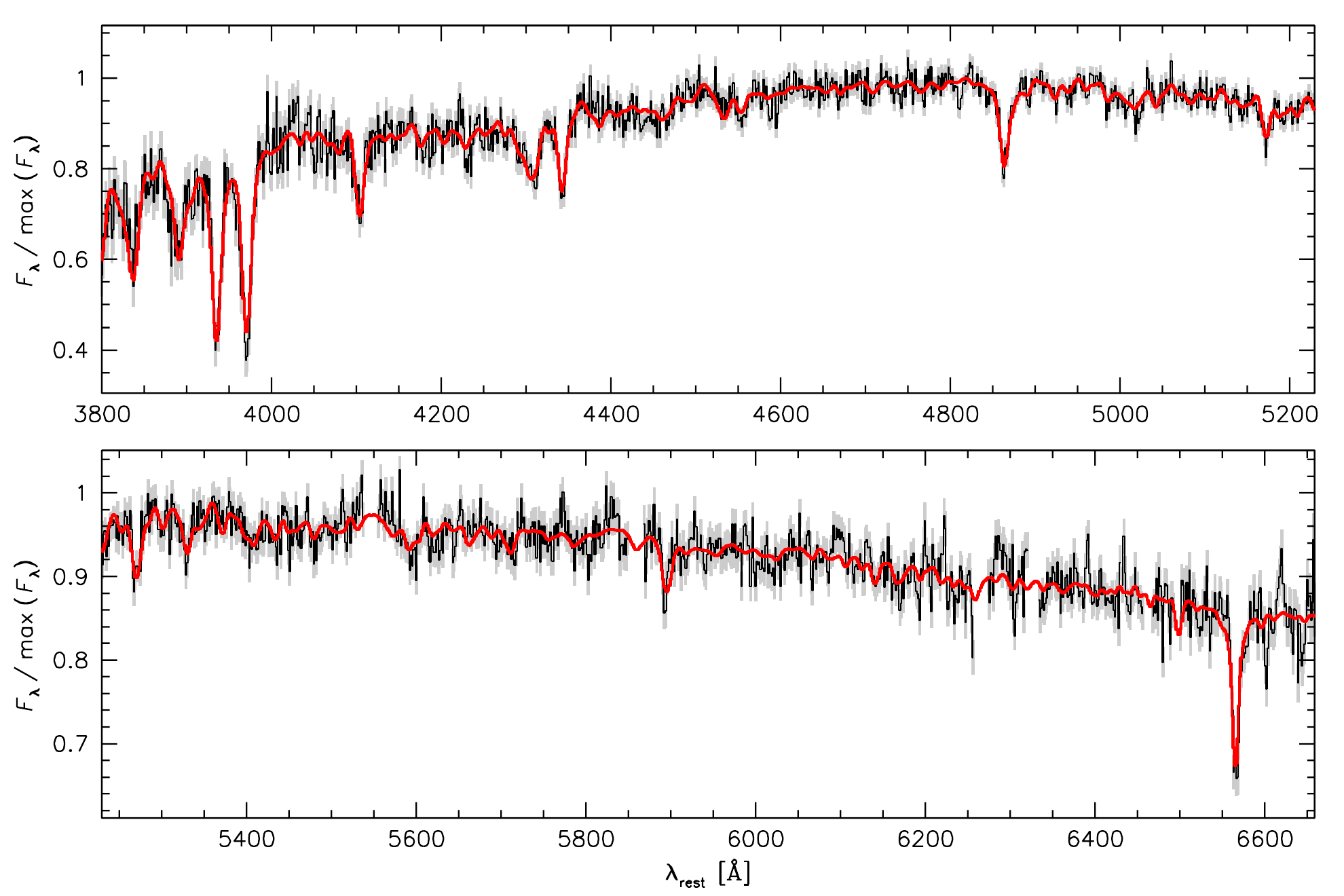}
  \end{center}
\vspace{-0.2cm}
    \caption{
Combined Keck/LRIS spectrum of the 11 GCs, weighted by the S/N ratio.
Errors are shown in grey. The best-fitting stellar population
synthesis model is shown in red. This model
has  an age of $9.3^{+1.3}_{-1.2}$\,Gyr,
${\rm [Fe/H]}=-1.35\pm 0.12$, and  ${\rm [Mg/Fe]}=0.16\pm
0.17$. The age is a lower limit, as it does not take the
possible presence of blue horizontal branch stars into account.
}\vspace{0.2cm}
\label{spec.fig}
\end{figure*}

We modeled
the LRIS-blue spectra with the most recent version
of the \texttt{alf} code ({Conroy} \& {van Dokkum} 2012; {Conroy} {et~al.} 2018).
% ( ), which cover a wide
%metallicity range ($-1.5 \lesssim {\rm [Fe/H]}\lesssim 0.3$).
%The empirical stellar library is described in
% () and the models, including fits to the
%integrated light spectra of Milky Way
%globular clusters, are described in {Conroy} {et~al.} (2018). 
%We find that the fits to the individual GCs 
%are consistent with each other but have large
%uncertainties.
To improve the constraints on the stellar population parameters
we stacked the 11 GC spectra, weighting by
the S/N ratio.
%This approach obviously removes any sensitivity to differences between
%the 11 clusters, but any
%variation is likely small given the observed color spread
%of $\sigma_{V-I}=0.039$.
The stacked spectrum is shown in Fig.\ \ref{spec.fig}.
The S/N ratio ranges from
$\approx 12$\,pix$^{-1}$ at $\lambda=3800$\,\AA\
to $\approx 55$\,pix$^{-1}$ at $\lambda=5400$\,\AA\ (with
$1.5$\,\AA\,pix$^{-1}$). 
The best fitting model, shown in red, has
${\rm [Fe/H]}=-1.35\pm 0.12$,  ${\rm [Mg/Fe]}=0.16\pm
0.17$, and ${\rm age}=9.3^{+1.3}_{-1.2}$\,Gyr.
The mass-to-light ratio is $M/L_V=1.8\pm 0.2$.
The errors were determined using an MCMC fitting technique,
as described in {Conroy} \& {van Dokkum} (2012).

We conclude that the objects are old and metal poor. This likely
applies to the entire system: the scatter in the $V_{606}-I_{814}$
colors of the GCs is very small, and their average color is
consistent with that of
the diffuse galaxy light: $\langle{}V_{606}-I_{814}\rangle_{\rm gc}=0.36
\pm{}0.02$ and
$(V_{606}-I_{814})_{\rm gal}=0.37\pm 0.05$.
%\footnote{The colors
%are also identical (within the errors)
%to those of both the GCs and the diffuse
%light in the Coma UDGs DFX1 and Dragonfly~44
%({van Dokkum} {et~al.} 2017).}

The $\alpha-$enhancement appears to be low, but typical values for
globular clusters ($0.3-0.5$) are only $1-2\sigma$ removed from the
best fit. Importantly, the
age (and also the $M/L$ ratio)
should be regarded as lower limits, due to the possible effects
of blue horizontal branch (BHB) stars. As discussed in, e.g.,
{Schiavon} (2007) and {Conroy} {et~al.} (2018) the presence of BHB stars reduces
the ages that are derived from integrated-light spectra.
%In {Conroy} {et~al.} (2018) we show explicitly that the integrated-light ages of
%metal-poor Galactic GCs underestimate the true age by 2--5 Gyr.
The average spectrum of the 11 \blob\ GCs is
similar to the integrated-light spectra of Galactic GCs with 
${\rm [Fe/H]}\sim -1.4$ and ages of $\sim 12$\,Gyr
(see {Mar{\'{\i}}n-Franch} {et~al.} 2009).

%The mass-to-light ratio is $M/L_V=1.8\pm 0.2$, for a  ()
%IMF. This implies a mean mass of the GCs of
%$M
%\approx 7\times 10^5$\,\msun. The brightest cluster, GC-73, has a mass
%of $M \approx 1.6 \times 10^6$\,\msun, approximately 1\,\% of the
%stellar mass of \blob.

% logage = 0.97 +/- 0.058
% [Fe/H] = -1.35+/-0.12
% [Mg/Fe] = 0.16 +/-0.17

\section{Discussion}

We analyzed the population of globular clusters
associated with the UDG \blob.
Superficially the galaxy resembles many other UDGs.
For example,
the morphology of the diffuse light and the fraction of the light
that is in GCs are similar
to the well-studied UDG Dragonfly~17 in the Coma cluster
({van Dokkum} {et~al.} 2015; {Peng} \& {Lim} 2016; {Beasley} \& {Trujillo} 2016). The stellar populations
are also similar; the $V_{606}-I_{814}$ colors 
are identical within the errors
to those of Dragonfly~44
({van Dokkum} {et~al.} 2017), and {Gu} {et~al.} (2017)
report ages and metallicities for three Coma UDGs that are consistent
with what we find here.
%\footnote{The colors
%are also identical (within the errors)
%to those of both the GCs and the diffuse
%light in the Coma UDGs DFX1 and Dragonfly~44
%({van Dokkum} {et~al.} 2017).}
%
A generic explanation for such diffuse, globular cluster-rich
systems may be that they are ``failed''
galaxies, in which star formation terminated 
shortly after the
metal-poor GCs appeared
and before a metal-rich component began to form. This naturally
explains their specific frequencies and uniform stellar populations,
and is qualitatively consistent with the observation that $S_N$
in dwarf galaxies is much higher when only metal-poor
stars are considered (e.g., {Larsen} {et~al.} 2014).

\blob\ is also very {\em different} from other UDGs (and indeed
all other known galaxies), in two distinct ways
that may be related to one another. First, the luminosity function
of the GCs has a narrow peak at $M_{V,606}\approx -9.1$ 
(Fig.~\ref{lf.fig}). This
is remarkable as the canonical value of $M_V\approx -7.5$
was thought to be universal, with only $\sim 0.2$\,mag
variation between galaxies
(see {Rejkuba} 2012). 
The origin of 
this unusual luminosity function is unknown; it
could be related to enhanced hierarchical
merging of lower mass clusters
(S.\ Trujillo-Gomez et al., in prep.).
The sizes and ellipticities of the GCs are different too, but this
may not be very fundamental. Since $\rho\propto{}Mr_h^{-3}$ the
GCs are a factor of $\sim{}2$ less dense than is typical. However,
their virial velocities are a factor of $\sqrt{2}$ higher, which
means that their kinetic energy densities 
$e_{\rm kin}\sim P\propto\rho{}v^2$ are similar.
Therefore, the same gas pressures were needed to form these
clusters as those that led to the formation of typical Galactic
GCs (see {Elmegreen} \& {Efremov} 1997). 
The higher ellipticities may simply reflect the initial angular
momentum of the GCs; as $t_{\rm r} \propto \sqrt{M}r_h^{1.5}$ the
relaxation times are a factor of $\sim 5$ longer than in typical
Milky Way GCs. We note that the effects of the external
gravitational potential on the structure of the GCs
are likely weak, due to the lack of dark
matter in \blob\ and the high masses of the clusters
(see, e.g., Goodwin 1997; Miholics et al.\ 2016).

The second difference is that the galaxy has no (or very little) dark
matter (see \natpap). This stands in stark contrast to cluster UDGs
(see {Beasley} {et~al.} 2016; {van Dokkum} {et~al.} 2016; {Mowla} {et~al.} 2017), and is inconsistent with
the idea that the old, metal-poor globular cluster systems
of galaxies are always closely connected to their
dark matter halos. Specifically,
previous studies
found that the ratio between the total mass in GCs and the
total (dark + baryonic) mass of galaxies is
remarkably constant, with $M_{\rm gc}^{\rm
tot}\approx{}3\times{}10^{-5}\,M_{\rm gal}^{\rm tot}$
(Blakeslee et al.\ 1997; {Harris} et al.\ 2015; {Forbes} {et~al.} 2016).
%UDGs in Coma and Virgo appear to follow this same
%relation ({Beasley} {et~al.} 2016; {Harris} {et~al.} 2017; {van Dokkum} {et~al.} 2017). 
%
%However, \blob is an extreme outlier.
Taking $M/L_V\approx 2$ (\S\,5) we find
$\approx{}9\times{}10^6$\,\msun\ for the total mass of the
globular clusters in \blob, and in \natpap\ we derived a
90\,\% upper limit of
$<3.4\times 10^8$\,\msun\ for its total galaxy mass.
Therefore, the mass in the GC system is $\gtrsim 3$\,\% of the
mass of the galaxy, a factor of $\sim 1000$ higher than the
Harris et al.\ value.
The existence of \blob\ suggests that the approximately linear
correlation
between GC system mass and total galaxy mass is not the result of 
a fundamental relation between the 
formation of metal-poor globular clusters and the
properties of dark matter halos
(as had been suggested by, e.g., {Spitler} \& {Forbes} 2009; {Trenti}, {Padoan}, \& {Jimenez} 2015; {Boylan-Kolchin} 2017).
Instead, the correlation may be
a by-product of other relations, with globular cluster formation
ultimately a baryon-driven process
(see, e.g., {Kruijssen} 2015; {Mandelker} {et~al.} 2017).
%spitler:09

Taking these ideas one step further, perhaps a key aspect of forming a UDG
-- or at least UDGs with many GCs -- is, paradoxically, the presence of
very dense gas at high redshift. After a short period of very intense
star formation the gas was blown out, possibly by supernova
(or black hole) feedback from the forming clumps themselves 
(e.g., {Calura} {et~al.} 2015). If the gas contained most of the mass
in the central regions of the forming galaxy this event may have led
to the extreme puffing up of the inner few kpc
(see also {Di Cintio} {et~al.} 2017; {Chan} {et~al.} 2017). The gas never returned,
either because the galaxy ended up in a cluster (Dragonfly 17) or
because it had very low mass (\blob). In this context having a massive
dark matter halo is not a central aspect of UDGs, but one of several
ways to reach sufficiently high gas densities for efficient
globular cluster formation at early times.

Of course, all this is speculation; also, this description of
events does not address the origin of
$\sim 10^{8-9}$\,\msun\ of extremely dense gas without a dark matter
halo. In this context,
an important unanswered question is whether
\blob\ is a ``pathological'' galaxy
that is the result of a rare set of circumstances
or representative of a class of similar objects. There are several
galaxies in our Cycle 23 {\em HST} program that superficially
resemble it, although none has quite as many luminous star 
clusters. %We are in the process of obtaining spectroscopy for these
%objects. 
\blob-like objects may have been more common in the past,
as large galaxies without dark matter lead a tenuous existence; in
clusters and massive groups they are easily destroyed, donating their
star clusters to the intracluster population of GCs and
ultra compact dwarfs (UCDs).
We note that progenitors  of galaxies like \blob\
could readily be identified in JWST observations
if its luminous GCs did indeed form within $\sim 10^8$\,yr
of each other in a dense region.

Finally, we briefly discuss whether the compact
objects in \blob\ should be considered globular clusters at all.
In terms of
their  average luminosity and size they are
intermediate between GCs and UCDs
(see, e.g., {Brodie} {et~al.} 2011), and
this question
hinges on whether we focus on the
population or on individual objects: the population characteristics
are unprecedented, but
for each individual object in \blob\ a match can be found
among the thousands of GCs with measured sizes and luminosities
in other galaxies
(e.g., {Larsen} {et~al.} 2001; {Barmby} {et~al.} 2007).
Intriguingly, in terms
of their sizes, flattening, stellar populations,
and luminosities the 11 compact
star clusters are remarkably similar to $\omega$\,Centauri --
an object whose nature has been the topic of decades of debate
(see, e.g., {Norris} \& {Da Costa} 1995).
%norris:95

%These galaxies were
%all found in low surface brightness imaging
%with the Dragonfly Telephoto Array,
%but we note that \blob\ could also have been identified by its
%globular cluster population. Searches for overdensities of unresolved
%objects in narrow color and magnitude ranges in wide field
%imaging data may turn up more examples of these enigmatic galaxies.

%  5.3 Formation scenarios
%    5.3.1 heating due to fly by?  (nature paper)
%    5.3.2 destruction; these are left overs?  
%       similar color -> but what mechanism?
%        Binney & Tremained life time estimate  ;
%       estimate mass of GCs too
%
%  5.4  Is DF2 unique ?
%
%    probably not!
%[ show DF2, DF4, 4258-DF4, all with zooms ]
%   nice thing is that they're accessible with
%     spectroscopy

\acknowledgements{Support from {\em HST} grant HST-GO-14644
and NSF grants AST-1312376, AST-1515084, 
AST-1518294, and AST-1613582
is gratefully acknowledged.
JMDK gratefully acknowledges funding from the German Research Foundation (DFG) in the form of an Emmy Noether Research Group (grant number KR4801/1-1) and from the European Research Council (ERC) under the European Union's Horizon 2020 research and innovation programme via the ERC Starting Grant MUSTANG (grant agreement number 714907).
AJR is a Research Corporation for Science Advancement Cottrell Scholar.
}

%% --------------------------------------------------------------------
%% Wed Jan 31 22:01:55 2018
%%   This file was generated automagically from the files
%%   blob_1052.bbl and blob_1052.tex using
%%     nat2jour.pl
%%   This file should accompany blob_1052-aas.tex.
%% --------------------------------------------------------------------

\end{document}